\documentclass[iop]{emulateapj}

\newcommand{\HII}{{\ion{H}{2}}}
\newcommand{\HeII}{{\ion{He}{2}}}

\newcommand{\OII}{[{\ion{O}{2}}]}

\newcommand{\OIIIHb}{[{\ion{O}{3}}]/H$\beta$}
\def\ratioR23{([\ion{O}{2}]~$\lambda$3727 +[\ion{O}{3}]~$\lambda\lambda$4959,5007)/H$\beta$}
\def\R23{${\rm R}_{23}$}

\newcommand{\Msun}{${\rm M}_{\odot}$}

\newcommand{\NII}{[{\ion{N}{2}}]}
\newcommand{\OIIIOII}{[\ion{O}{3}]/[\ion{O}{2}]}

\newcommand{\SIIl}{[\ion{S}{2}]~$\lambda \lambda$6717,31}

\newcommand{\OH}{$\log({\rm O/H})+12$}

\newcommand{\NIIHa}{[\ion{N}{2}]/H$\alpha$}
\newcommand{\SIIHa}{[\ion{S}{2}]/H$\alpha$}
\newcommand{\OIHa}{[\ion{O}{1}]/H$\alpha$}

\newcommand{\SII}{[{\ion{S}{2}}]}

\def\O4363{[{\ion{O}{3}}]~$\lambda$4363}
\newcommand{\OIII}{[{\ion{O}{3}}]}

\newcommand{\OI}{[{\ion{O}{1}}]~$\lambda$6300}
\newcommand{\Ha}{{H$\alpha$}}

\def\L60{L$_{60}$}

\shorttitle{}
\shortauthors{}

\begin{document}

\title{Theoretical Evolution of Optical Strong Lines across Cosmic Time}

\author{Lisa J. Kewley}
\affil{Australian National University}
\affil{University of Hawaii}
\email{kewley@mso.anu.edu.au}

\author{Michael A. Dopita}
\affil{Australian National University}
\affil{King Abdulaziz University}

\author{Claus Leitherer}
\affil{STScI}

\author{Romeel Dav\'{e}}
\affil{University of Arizona}
\affil{University of the Western Cape}
\affil{South African Astronomical Observatory}
\affil{African Institute for Mathematical Science}

\author{Tiantian Yuan}
\affil{University of Hawaii}

\author{Mark Allen}
\affil{University of Strasbourg}

\author{Brent Groves}
\affil{Leiden University}

\and

\author{Ralph Sutherland}
\affil{Australian National University}

\begin{abstract}
We use the chemical evolution predictions of cosmological hydrodynamic simulations with our latest theoretical stellar population synthesis, photoionization and shock models to predict the strong line evolution of ensembles of galaxies from $z=3$ to the present day.  In this paper, we focus on the brightest optical emission-line ratios, \NIIHa\ and \OIIIHb.  We use the optical diagnostic Baldwin-Phillips-Terlevich (BPT) diagram as a tool for investigating the spectral properties of ensembles of active galaxies.  We use four redshift windows chosen to exploit new near-infrared multi-object spectrographs.   We predict how the BPT diagram will appear in these four redshift windows given different sets of assumptions.  We show that the position of star-forming galaxies on the BPT diagram traces the ISM conditions and radiation field in galaxies at a given redshift.   Galaxies containing AGN form a mixing sequence with purely star-forming galaxies.  This mixing sequence may change dramatically with cosmic time, due to the metallicity sensitivity of the optical emission-lines.  Furthermore, the position of the mixing sequence may probe metallicity gradients in galaxies as a function of redshift, depending on the size of the AGN narrow line region.  We apply our latest slow shock models for gas shocked by galactic-scale winds.  We show that at high redshift, galactic wind shocks are clearly separated from AGN in line ratio space.  Instead, shocks from galactic winds mimic high metallicity starburst galaxies.  We discuss our models in the context of future large near-infrared spectroscopic surveys.
\end{abstract}


\section{Introduction}
Understanding how galaxies formed and evolved is one of the primary drivers of modern astronomical research.  Measuring the fundamental properties of galaxies as a function of cosmic time is now possible by combining spectroscopy from the world's largest telescopes with state-of-the-art theoretical simulations.  



The collisionally excited emission-lines provide key diagnostics of the gas-phase chemical abundance, the ionization state of the gas, the dust extinction, and the ionizing power source of the galaxy.  \citet{Baldwin81} showed that a diagram of \NIIHa\ and \OIIIHb\ (commonly referred to as the BPT diagram) can be used to classify galaxies dominated by AGN from those dominated by star formation.  The extreme ultra-violet (EUV) ``hard" radiation field from the accretion disk of an AGN ionizes the \OIII\ and \NII\ lines, producing larger \OIIIHb\ and \NIIHa\ line ratios than usually seen in star-forming galaxies.     \citet{osterbrock85} and \citet[][hereafter VO87]{Veilleux87} derived the first semi-empirical AGN classification schemes for the BPT diagram based on a combination of observations and and photoionization models.   These diagnostics were refined by \citet{Kewley01b} using stellar population synthesis, photoionization and shock models.  

Star-forming galaxies form a tight sequence on the BPT diagram, known as the star-forming or \HII\ ``abundance sequence" \citep{Dopita86,Dopita00}.  The location of this abundance sequence probes (1) the spread in global metallicity of the observed star-forming galaxy population, (2) the stellar ionizing radiation field of the star-forming population, and (3) the conditions of the interstellar medium (ISM) surrounding the star-forming regions.   Therefore, the  \NIIHa\ versus \OIIIHb\ diagram may be used as a tool for investigating the metallicity, ISM, and ionizing radiation field in star-forming galaxies as a function of cosmic time,  independent of the large systematic errors that plague the chemical abundance scale \citep{Kewley08,Bresolin09,Kudritzki12}.  

The gas-phase metallicity has a critical influence on the location of both star-forming galaxies and AGN on the BPT diagram.  Power-law AGN models show that the position and spread of the AGN region on the BPT diagram traces the metallicity in the extended narrow-line region of AGN \citep{Groves04b}.  Both theory and observations indicate that the mean metallicity of galaxies rises with cosmic time as galaxies undergo successive generations of star formation \citep[e.g.,][]{Nagamine01,deLucia04,Kobulnicky04,Kobayashi07,Maiolino08,Dave11b,Yuan12,Zahid13}.    The metallicity evolution of galaxies will therefore change the position of galaxies on the BPT diagram as a function of redshift.  Because optical classification of starburst and AGN is based on either empirical fits to local galaxies, or theoretical models developed for local galaxies,  current BPT classification methods may not be applicable at high redshift.  


Testing BPT classification methods at intermediate or high redshift has been difficult in the past.  At $z>0.4$, the \NII\ and \Ha\ lines are redshifted into the near-infrared.  With single-slit near-infrared spectroscopy, \NIIHa\ and \OIIIHb\ ratios have now been observed for small numbers of individual galaxies at high redshift \citep{Teplitz00,Finkelstein09,Hainline09,Bian10,Rigby11,Yabe12}.   Stacked \NIIHa\ and \OIIIHb\ ratios of large numbers of galaxies have also been measured \citep{Erb06}, and the first large NIR spectroscopic surveys are now being conducted \citep[e.g.,][]{Trump13}.  
 
The majority of galaxies at $z>0.4$ show an offset towards larger \NIIHa\ and \OIIIHb\ ratios compared with local galaxies.  This offset may be caused by a higher ($2\times$) ionization parameter in high redshift galaxies.  A larger ionization parameter may be produced by high nebular electron densities, a higher rate of star formation, a top-heavy initial mass-function (IMF), a high volume filling factor, and a large escape fraction of UV photons \citep{Brinchmann08b}.   \citet{Lehnert09} suggest that the offset results from high gas densities and pressures that are similar to the most intense nearby SF regions locally,  but spread over scales of 10-20~kpc in high redshift galaxies.  Higher electron densities and ionization parameters have been measured for several high redshift galaxies \citep{Hainline09,Bian10,Liu08}, but a large electron density by itself cannot explain the offset in all cases \citep[see][]{Rigby11}.   \citet{Groves06} use photoionization models to show that the emission-line ratios in some high redshift galaxies could be explained by a combination of starburst and AGN activity.   A similar conclusion was reached by \citet{Trump11} after combining HST spectra with Chandra X-ray data for a sample of galaxies at $z\sim 2$. Concurrent star-formation and AGN activity has also been found in some high redshift lensed galaxies \citep{Wright10}.

The astronomical community is now on the cusp of obtaining the BPT diagnostic emission lines for large samples of $z>0.4$ galaxies for the first time, thanks to new near-infrared multi-object spectrographs, such as  MOSFIRE on Keck \citep{Mclean10}, FMOS on Subaru \citep{Kimura10}, MMIRS on Magellan \citep{Mcleod04}, and FLAMINGOS~II on Gemini \citep{Eikenberry08}.  
To interpret these spectra, a theoretical understanding of how the key BPT features may change at high redshift is needed.  Such an understanding is essential for separating samples of star-forming galaxies from AGN at high redshift, and for tracking fundamental physical properties of the active galaxy population with redshift.   

In this paper, we combine the predictions of  cosmological hydrodynamic simulations models with stellar evolution, photoionization and shock models to predict how the BPT diagram may change with redshift.  In Section~\ref{cosmological_hydrodynamic}, we describe the theoretical chemical evolution predictions used.
Our simulations for the star-forming galaxy abundance sequence are given in Section~\ref{starburst_models}.  In Section~\ref{mixing_sequence}, we model how the the line ratios of galaxies containing AGN may evolve with redshift.  Section~\ref{Cosmic_BPT} gives theoretical predictions of the position of the BPT diagram within specific redshift windows.
We investigate contamination from shocks, and we discuss the limitations of this work.  Our conclusions are presented in Section~\ref{conclusions}.
Throughout this paper, we adopt the flat $\Lambda$-dominated cosmology as measured by the 7 year WMAP experiment \citep[$h=0.72$, $\Omega_{m}=0.29$;][]{Komatsu11}.

\section{Chemical Evolution of Active Galaxies}\label{cosmological_hydrodynamic}


We utilize the chemical evolution estimates from \citet{Dave11a,Dave11b}.  The \citet{Dave11a} models use the GADGET-2 N-body $+$ Smoothed Particle Hydrodynamic code \citep{Springel05} in a $\Lambda{\rm CDM}$ cosmology.  This code incorporates gas cooling and heating processes, including the effects of metal line cooling \citep{Oppenheimer06}.  Density-driven star formation is calculated using a Schmidt law \citep{Schmidt59}.   Chemical enrichment from Type II supernovae, Type I supernovae, and AGB stars is included.  Dave et al. use a Monte Carlo approach to model galactic outflows, where the mass loss due to outflows is related to the star formation rate and a variable mass loading factor.

We use the predicted evolution in the gas-phase chemical abundance for star-forming galaxies with stellar mass $ {\rm M}_{*} >  10^9 
{M}_{\odot} $ across $0<z<3$.  We parameterize the relative change in chemical abundances, $\Delta(\log{\rm O/H})$, from $z=0$ to an arbitrary redshift $z$ by a 3rd order polynomial:

\begin{equation}
\Delta(\log{\rm O/H}) =  -0.0013 - 0.2287 z +0.0627 z ^2 -0.0070 z^3 \label{eqn_metal_z}.
\end{equation}

The $1 \sigma$ error about equation~\ref{eqn_metal_z} is $\pm 0.1$~dex at $z=0$, falling to $\pm 0.05$~dex at z=3.  

Equation~\ref{eqn_metal_z} assumes that there is no mass dependence in the chemical evolution for $ {\rm M}_{*} >  10^9 {M}_{\odot} $ and over $0<z<3$.   We show in \citet{Yuan12} that equation~\ref{eqn_metal_z} fits the slope of the current metallicity history of galaxies for $ {\rm M}_{*} >  10^9 {M}_{\odot} $ to within the observational errors.    In future, when the chemical enrichment history of galaxies as a function of stellar mass is understood, a mass term could be included in equation~\ref{eqn_metal_z}, if needed.

\section{The Star-forming Abundance Sequence} \label{starburst_models}

We combine stellar evolutionary synthesis models with our MAPPINGS III photoionization models to generate theoretical limits to the 
expected ionizing radiation field as a function of redshift.  
We have tested this combination of stellar population synthesis and photoionization models extensively for local star forming galaxies \citep[e.g.,][]{Kewley01a,Levesque10}.
Zero-age and 1~Myr old models are able to reproduce the \NIIHa\ and \OIIIHb\ line ratios in the majority of star-forming galaxies.  However, the \SIIHa\ and \OIHa\ line ratios require a harder ionizing radiation field than is available in current stellar population synthesis models \citep{Levesque10}.   This situation may be resolved when the effects of stellar rotation are incorporated into the stellar evolutionary tracks used by the population synthesis models.  Initial investigations into the effect of stellar rotation on the ionizing radiation field at solar metallicity is promising \citep{Levesque12}.  Until the full set of stellar tracks with rotation become available, we limit our analysis of BPT evolution to the \NIIHa\ and \OIIIHb\ emission-line ratios which can already be reproduced using current stellar population synthesis models.

\subsection{The Local Star-forming Abundance Sequence} \label{abundance_sequence}

The Starburst99 (SB99) models that we use are described in detail in \citet{Levesque10} and \citet{Nicholls12}.  Briefly, we apply 
a Salpeter IMF \citep{Salpeter55}  with an upper mass limit of 100 \Msun.   The choice of IMF makes negligible difference on the optical emission-line ratios used in this analysis.  We use the Pauldrach/Hillier model atmospheres, which employ the WMBASIC wind models of \citet{Pauldrach01} for younger ages when O stars dominate the luminosity ($<3$~Myr), and the CMFGEN  atmospheres from \citet{Hillier98} for later ages when Wolf-Rayet (W-R) stars dominate.  These stellar atmosphere models include the effects of metal opacities.   We use the Geneva group ``high" mass-loss evolutionary tracks \citep{Meynet94}.  These tracks include enhanced mass-loss rates that are applicable to low-luminosity W-R stars and can reproduce the blue-to-red supergiant ratios observed in the Magellanic Clouds \citep{Schaller92,Meynet93}.  Starburst99 generates a  synthetic FUV spectrum using isochrone synthesis \citep[e.g.,][]{Charlot91}, in which isochrones are fitted to the evolutionary tracks across different masses rather than discretely assigning stellar mass bins to specific tracks.   We use the zero-age instantaneous burst models because these models provide the best fit to the SDSS star-forming galaxy sequence at $z\sim0$ \citep{Dopita13}.   To match the nebular metallicities of our photoionization code, we interpolate between the STARBURST99 model grids as a function of metallicity.     

We use our Mappings IV photoionization code \citep{Binette85,Sutherland93,Dopita13} to model the interstellar medium surrounding the SB99 ionizing radiation field.   We assume the solar abundance set of \citet{Asplund05}.  This abundance set includes revised solar abundances for key elements, including oxygen and carbon.  As in \citet{Levesque10}, the $\alpha$-element abundances are assumed to scale linearly with metallicity, with the exception of Helium and Nitrogen.  For helium, we include the stellar yield in addition to the primordial abundance from \citet{Pagel92}.  For nitrogen, we assume primary and secondary nucleosynthetic components as measured by
\citet{Mouhcine02} and \citet{Kennicutt03}.  The resulting ${\rm N/H}$ ratio is parameterized in \citet{Groves04a}.
Metals are depleted out of gas phase and onto dust grains.  The dust depletion factors are given in \citet{Groves06}, and are based on \citet{Kimura03}, who examined the metal absorption along several lines of sight within the local interstellar cloud. 
 
Mappings uses either a plane parallel or spherical geometry with the ionization parameter defined at the initial edge of the nebula.    For a spherical geometry, an effective ionization parameter $q$ can be defined that takes into account the spherical divergence of radiation at the Stromgren radius $R_s$ \citep{Stromgren39}:

\begin{equation}
q_{eff}=\frac{Q_{{\rm H}^{0}}}{4 \pi {{R}_s}^2 n_H}  \label{q_spherical}
\end{equation}

where $Q_{{\rm H}^{0}}$ is the flux of ionizing photons above the Lyman limit.  If the thermal gas only occupies a fraction of the available volume, the ionization parameter can be defined in terms of a volume filling factor.   

The ionization parameter $q$ has units of velocity (cm/s) and can be thought of as the maximum velocity ionization front that an ionizing radiation field is able to drive through a nebula.  This dimensional ionization parameter is related to the dimensionless ionization parameter $U$ through the identity $U\equiv q/c$.  The dimensional ionization parameter is typically $ -3.2 < \log U < -2.9$ for local \HII\ regions \citep{Dopita00} and star-forming galaxies \citep{Moustakas06,Moustakas10}.  In practice, all models with a similar effective ionization parameter produce very similar spectra, assuming all other parameters are held constant \citep[e.g.,][]{Dopita00}.

To minimize small uncertainties produced by particular geometries, we calculate spherical models in which $q$ is determined at the inner radius.   The average ionization parameter is lower than this initial value, and is dependent on the ionization parameter of the initial radius.    Models were run with pressure ${\rm P}/k = 10^{5.5}\,{\rm cm}^{-3}~{\rm K}$, where $k$ is the Boltzmann constant.  In an ionized nebula, electron temperatures are $\sim 10^4$~K, yielding a density of $10-30$~${\rm cm}^{-3}$, typical of giant extragalactic \HII\ regions.  Detailed photoionization, excitation, and recombination are calculated at increments (step size $0.03$) throughout the nebula.  The model completes when the hydrogen gas is fully recombined.  A full description of the models, including geometry, is given in \citet{Lopez12}.

Unlike the \citet{Kewley02} models, our current models include a sophisticated treatment of dust, including the effects of absorption, grain charging, radiation pressure, and photoelectric heating of the small grains \citep{Groves04a}.  The latest version of Mappings incorporates a Kappa temperature distribution that is a more realistic representation of the electron temperature distribution in a turbulent ISM than a Stefan-Boltzmann distribution \citep{Nicholls12}.     

\begin{figure*}[!t]
\epsscale{1.3}
\plotone{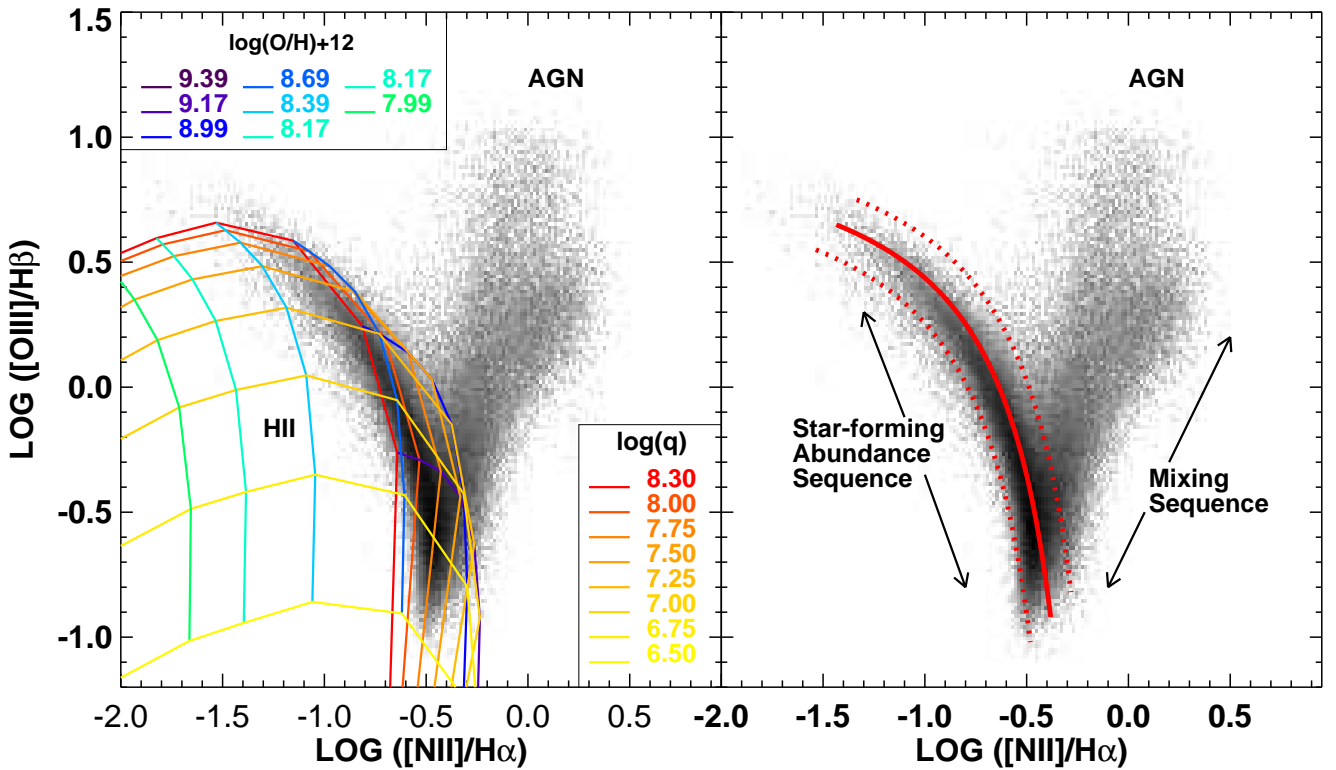}
\caption[Kappa_BPT.ps]{The \NIIHa\ versus \OIIIHb\ optical diagnostic diagram for the Sloan Digital Sky Survey galaxies analyzed by \citet{Kewley06a}.   Left: The colored curves show our new theoretical stellar population synthesis and photoionization model grid for star-forming galaxies based on a $\kappa$ electron temperature distribution.  Right: The red solid curve shows the mean star-forming sequence for local galaxies.  The shape of the red solid curve is defined by our theoretical photoionization models, while the position is defined by the best-fit to the SDSS galaxies.  The $\pm 0.1$ dex curves (dashed lines) represent our model errors and contain 91\% of the SDSS star-forming galaxies.
\label{Kappa_BPT}}
\end{figure*}

Figure~\ref{Kappa_BPT} (left panel) shows our model grid in comparison to the local star-forming galaxy sequence from the Sloan Digital Sky Survey (SDSS) from \citet{Kewley06a}.  The SDSS galaxies were selected within the redshift range of $0.04 <z<0.1$ to minimize aperture effects and Malmquist bias \citep{Kewley04}.  

We fit an equation of the form $\log [{\rm OIII}]/{\rm H}\beta =\frac{a}{\log ( [{\rm NII}]/{\rm H}\alpha)+b}+c$ to our model grid, where $a,b,c$ are constants.  This polynomial form was chosen for consistency with previous fits to theoretical model grids, including the hard ionizing radiation field from the P\`{e}gase stellar population synthesis models \citep{Kewley01a,Kauffmann03b}.    According to our models, the mean position of a galaxy along the local star-forming sequence is 

 \begin{equation}
\log(\frac{[OIII]}{{\rm H}\beta}) = \frac{0.61}{ \log({NII}/{\rm H}\alpha) +0.08 }+1.1   \label{SDSS_sequence}
\end{equation}

 In Figure~1 (right panel), we show the polynomial fit from equation~\ref{SDSS_sequence}.  This polynomial represents the mean local star-forming abundance sequence with model estimated errors of $\pm 0.1$~dex in both the \NIIHa\ and \OIIIHb\ directions.   The upper and lower bounds of $\pm 0.1$~dex (dashed lines) encompass  91\% of the SDSS star-forming galaxies.

The \NIIHa\ and \OIIIHb\ line ratios along equation~\ref{SDSS_sequence} are related to metallicity through the following equation

\begin{equation}
 12 + \log{\rm O/H} =  8.97 -  0.32 x, \label{Z_SDSS_sequence}
\end{equation}

where  $x = \log \frac{[OIII]/H\beta}{[NII]/H\alpha}$ and metallicity is defined on the \citet{Kewley02} metallicity scale.  We note that the following analysis depends only on the {\it relative} change in metallicity along the local starburst abundance sequence (equation~\ref{SDSS_sequence}).   Different metallicity diagnostics have systematic offsets in their absolute abundance scale but relative metallicity measurements are conserved to within $\pm 0.03$~dex on average \citep{Kewley08}.   Thanks to this conservation of relative metallicities, the relative change in position along equation~\ref{SDSS_sequence} is independent of the metallicity calibration used.

\subsection{Factors that affect the Star-Forming Abundance Sequence}

Individual galaxies that lie on the BPT diagram at high redshift are likely to evolve off the BPT diagram before $z=0$.  Thus, an observed change in the star-forming abundance sequence will track the change in intrinsic galaxy properties across the star forming galaxy population at different epochs.  

The position of our theoretical star-forming galaxy abundance sequence is determined by: (1) the shape of the ionizing radiation field, (2) the geometrical distribution of gas with respect to the ionizing sources, (3) the metallicity range, and (4) the electron density (pressure) of the gas.   We discuss the effect of changing each of these quantities below.

\subsubsection{Shape of the Ionizing Radiation Field}

The stellar ionizing radiation field may change with redshift as a result of a change in the fraction of ionizing photons produced by the young stellar population.  
In a pure star forming galaxy, the hardness of the ionizing radiation field is related to the slope of the IMF, the age of the stellar population and the metallicity of the galaxy.   A stellar population with a shallow initial mass function produces a hard ionizing radiation field, but there is no solid evidence for a change in IMF with redshift  \citep[see][]{Bastian10,Greggio12}.

 The stellar population age is directly related to the shape of the ionizing radiation field.  Hard ionizing radiation fields can be produced at $\sim 3-5$~Myr when the stellar population may be dominated by Wolf-Rayet stars \citep[e.g.,][]{Schaerer96,Kehrig08}.   Broad \HeII~$\lambda 1640$ emission has been observed in stacked spectra of Lyman Break Galaxies \citep{Shapley03} and in some individual high redshift gravitationally lensed galaxies \citep{Cabanac08,Dessauges10}.  The  broad \HeII\ feature has been attributed to a significant contribution from O and Wolf-Rayet stars to the ionizing EUV radiation field at low metallicity \citep{Brinchmann08a}.  We note that radiative shocks can produce a narrow, nebular  \HeII\ feature \citep{Dopita11,Lagos12} which may be blended with the broad component produced in stellar atmospheres of luminous stars.  Such blending could be difficult to distinguish at high redshift.

A hard ionizing radiation field has been linked with low metallicity in star forming galaxies \citep[e.g.,][]{Campbell86,Galliano05,Madden06,Hunt10,Levesque10}.   There are several potential reasons for the correlation between metallicity and the hardness of the ionizing radiation field: 

(1) High energy photons produced by metal-rich stars are absorbed by metals in the stellar atmosphere, known as metal blanketing.  The preferential absorption of high energy photons yields a softer ionizing radiation field  \citep[e.g.,][]{Gonzalez05}. 

(2) The Hayashi track shifts to hotter effective temperatures at low metallicities, enabling metal-rich massive stars to maintain a higher effective temperature compared with metal-poor stars of similar spectral types \citep{Elias85,Levesque06}.

(3) Low metallicities correspond to lower mass loss rates, allowing low metallicity stars to remain on the main sequence for longer timescales \citep{Meynet94,Maeder94}.

(4)   In isolated stars, rotational mixing causes heavy mass loss.  This mass loss produces bluer colors in the red supergiant phase, lowering the mass limit required for a star to enter the Wolf-Rayet phase \citep{Levesque12}.   Thus, a population of rotating massive stars will contain a larger fraction of hot, massive stars to contribute ionizing photons to the stellar radiation field.    This rotational hardening is a function of metallicity, with more significant hardening in metal-poor environments \citep{Leitherer08}.    

(5)  In binary stars, efficient mass-transfer can spin-up the rotation of the companion star, causing similar mixing effects as in rotating isolated stars \citep{deMink09,Eldridge12}.  Whether a rapidly rotating star can spin down depends on stellar winds, which are weaker at low metallicities due to metal opacity.   Rotation and binarity are not yet included in stellar population synthesis models for a range of metallicities.

A hard ionizing radiation field can also be produced by contamination from an active galactic nucleus or radiative shocks.  Slow shocks (100-200 km/s)  associated with galactic-scale winds have been observed both locally \citep{Rich10,Rich12} and at high redshift \citep{Yuan12}.  We address these two possibilities in Section~\ref{mixing_sequence} and Section~\ref{shocks}.

\begin{figure}[!h]
\epsscale{1.2}
\plotone{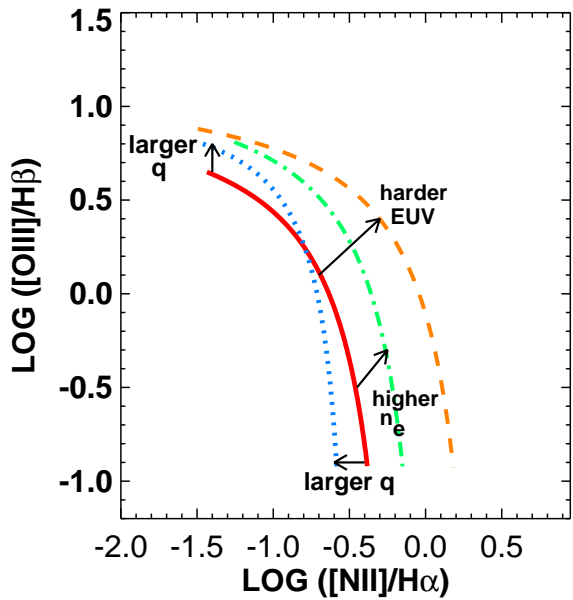}
\caption[BPT_illustration.ps]{An illustration of the effect of varying different galaxy parameters on the star-forming galaxy abundance sequence in the \NIIHa\ versus \OIIIHb\ diagnostic diagram.   The original SDSS star-forming galaxy sequence is well-fit by the red theoretical curve.  Raising the hardness of the ionizing radiation field (orange dashed line) moves the abundance sequence towards larger \NIIHa\ and \OIIIHb\ ratios.  A similar effect is seen when the electron density of the gas is raised (green dot-dashed line).   The relationship between ionization parameter, metallicity and the \NIIHa\ and \OIIIHb\ line ratios is more complex.  At high metallicities, raising the ionization parameter causes the \NIIHa\ ratio to become smaller, while \OIIIHb\ is largely unaffected.   At low metallicities, raising the ionization parameter raises the \OIIIHb\ ratio while \NIIHa\ is largely unaffected.
\label{BPT_illustration}}
\end{figure}

It is unclear whether high redshift star-forming galaxies have a harder radiation field than local galaxies at the same metallicity.   If a harder ionizing radiation field exists in high redshift galaxies, both the \OIIIHb\ and \NIIHa\ line ratios will be affected.  A larger fraction of photons with energies above the ionization potentials of \NII\ and \OIII\  will raise the \NIIHa\ and \OIIIHb\ line ratios because there are more photons available to ionize nitrogen and oxygen.  The \OIII\ emission-line is substantially more sensitive to the hardness of the EUV ionizing radiation field than \NII\ because the difference in ionization potentials is large (35eV   c.f.  14.5eV).  
In Figure~\ref{BPT_illustration}, we illustrate the effect of raising the hardness of the ionizing radiation field in galaxies along the local abundance sequence (orange dashed line).  The harder radiation field moves the entire abundance sequence above and to the right on the BPT diagram.  A harder ionizing radiation field could account for part (or all) of the large \NIIHa\ and \OIIIHb\ ratios seen at high redshift.

\subsubsection{Geometrical distribution of the gas}

The geometrical distribution of the gas around the ionizing source can be changed by shocked stellar winds.  Stellar winds clear ionized gas from the interior of an \HII\ region.  Because highly ionized species, such as \OIII, form preferentially at the inner radii of \HII\ regions, shocked stellar winds will lower the effective ionization parameter of the nebula \citep{Yeh12}.

Our theoretical photoionization models assume a uniform medium, where the geometrical distribution of the gas is approximated by a volume filling factor.  In reality, \HII\ regions may be clumpy and porous.  Ultracompact \HII\ regions in the Milky Way contain a porous interstellar medium \citep{Kurtz99,Kim01}.    Radio observations indicate that young \HII\ regions in low metallicity galaxies also have a clumpy and porous distribution of gas  \citep{Johnson09}.   A clumpy, porous medium allows some ionizing photons to escape the nebula without being absorbed by the interstellar gas.  In this scenario, the effective ionization parameter may be lowered, depending on the escape fraction and the optical thickness of the porous medium.

Our models are calculated for radiation-bounded \HII\ regions.  In the radiation-bounded scenario, the model finishes when hydrogen is completely recombined.  If \HII\ regions are density bounded (i.e. the \HII\ region density is sufficiently low that the stars can ionize the entire nebula), the \OIII\ ionization zone is likely to be largely unaffected, but the \NII\ excitation and Hydrogen recombination zones may be shortened.  Therefore, the \OIIIHb\ ratio may be larger, while the \NIIHa\ ratio will be similar to or smaller than observed in a radiation-bounded nebula.   Since the \OII\ zone is shorter in a density-bounded nebula, the \OIIIOII\ ratio becomes larger.  If radiation-bounded models are applied to such nebulae, then the larger \OIIIOII\ line ratio would be interpreted as a high ionization parameter.   

Detailed ionization parameter mapping of the \HII\ regions in nearby galaxies can constrain whether the \HII\ regions are radiation bounded or density bounded.   A mixture of radiation-bounded and density-bounded \HII\ regions have been observed in the local group \citep{Pellegrini12}.   \citet{Nakajima12} suggest that Lyman-$\alpha$ emitters at high redshift contain density-bounded \HII\ regions.  It is unclear whether density-bounded nebulae are common in normal star-forming galaxies, either locally or at high redshift.

The geometrical distribution issues described above can be considered in terms of an effective ionization parameter.  We discuss the effect of changing the ionization parameter on the \NIIHa\ and \OIIIHb\ line ratios in Section~\ref{q}.

\subsubsection{The Metallicity Range of Galaxies}

The spread in metallicity across a galaxy sample determines the length of the star-forming abundance sequence.  Galaxy samples that span a small range of metallicities occupy only a portion of the star-forming abundance sequence.  For example, interacting or merging galaxies typically have lower central metallicities due to large-scale gas infall 
\citep{Kewley06b,Kewley10,Rupke10b,Ellison10,Rich12,Scudder12}.   On the other hand, low metallicity galaxies, such as blue compact dwarfs, occupy the highest positions (lowest metallicities) on the star-forming abundance sequence \citep{Levesque10}. 

The metallicity range of high redshift galaxies is unknown.  Samples selected from rest-frame blue colors or the Lyman break may be missing a population of faint, low metallicity galaxies and a population of dusty, metal-rich star forming galaxies.   Gravitationally lensed samples probe fainter galaxy samples and a broader range of metallicities, within current instrumentation detection limits  \citep[see Figure~5 in ][for a comparison of current instrumentation detection limits for lensed and non-lensed samples]{Yuan12}.

\subsubsection{The Electron Density}

In an isobaric density distribution, the density is defined in terms of the ratio of the mean ISM pressure, $P$,  and mean electron temperature, $T_e$, through $n_e =\frac{ P }{ T_e k }$.  For an ionized gas, the electron temperature is $\sim 10^4$~K and the density is simply determined by the ISM pressure.

The SDSS abundance sequence is fit by our photoionization models with an electron density of $n_e = 10 - 10^2~{\rm cm^{-3}}$, typical of local \HII\ regions \citep{Osterbrock89}.  
\citet{Brinchmann08b} shows that the distance away from the star-forming abundance sequence correlates strongly with electron density; 
SDSS galaxies with larger electron densities ($n_e = 10^2~{\rm cm^{-3}}$) lie above and to the right of the mean SDSS star-forming abundance sequence.   Highly active star-forming galaxies, such as warm infrared galaxies and luminous infrared galaxies have dense ionized gas \citep[$10^2 -  10^3~{\rm cm^{-3}}$;][]{Kewley01b,Armus04} and lie above and to the right of the SDSS star-forming abundance sequence \citep{Yuan10}.

Many high redshift galaxies also have dense nebular gas  ($\sim 10^3~{\rm cm^{-3}}$) \citep[e.g.][]{Brinchmann08b,Hainline09,Bian10,Rigby11,Wuyts12} and galaxies with high \Ha\ surface brightness \citep{Lehnert09,LeTiran11}.   Figure~\ref{BPT_illustration} shows the effect of raising the electron density from $n_e = 10~{\rm cm^{-3}}$ (solid red line) to $n_e = 1000~{\rm cm^{-3}}$ (dot-dashed green line).  Both the \OIIIHb\ and \NIIHa\ ratios are larger at higher electron densities due to the increased rate of collisional excitation.   We conclude that high electron densities could account for at least part of the enhanced \NIIHa\ and \OIIIHb\ ratios observed in current samples of high redshift galaxies.

\subsubsection{Ionization Parameter}\label{q}

Much previous research into the differences between high redshift and local galaxy emission-line ratios has focused on the ionization parameter \citep{Brinchmann08b,Liu08,Hainline09}.   The ionization parameter is a useful tool for comparisons among \HII\ regions or galaxies because it can be readily measured using the ratio of high ionization to low ionization species of the same atom (such as \OIIIOII).  However, it is important to understand the fundamental physical properties that govern the ionization parameter.  The ionization parameter is determined by the hydrogen ionizing photon flux, and the pressure of the interstellar medium.

The number of hydrogen ionizing photons per unit area can be changed by either scaling the ionizing radiation field (i.e. raising or lowering the luminosity of the stellar population) or by modifying the shape (hardness) of the ionizing radiation field.   Because the ionization parameter, electron density and radiation field hardness are interrelated, isolating the cause(s) of a shift towards larger \NIIHa\ and/or \OIIIHb\ ratios at high redshift is non-trivial.  Furthermore, different geometrical gas distributions can mimic a higher or lower observed ionization parameter.  Several of these processes may contribute to the observed ionization parameter in star-forming galaxies.

Local galaxies with high measured or inferred ionization parameters include Wolf-Rayet galaxies \citep{Brinchmann08b}, strong \Ha\ emitter galaxies \citep{Shim12}, and low metallicity galaxies \citep{Kewley07}.    In Wolf-Rayet galaxies, the high ionization parameter is caused by the shape of the ionizing radiation field produced in Wolf-Rayet atmospheres with line blanketing \citep{Gonzalez02}.  Strong \Ha\ emitting galaxies may have high ionization parameters due to a high ISM pressure \citep{Brinchmann08b,Shim12}, and/or low metallicity.  
  
The ionization field can be scaled by raising the star formation rate of a stellar population.
High redshift galaxies typically have larger star formation rates than local galaxies \citep{Shapley05,Hainline09}.  \citet{Hainline09} suggest that these larger star formation rates may lead to a larger reservoir of ionizing photons, leading to a higher ionization parameter.

Alternatively, a large ionization parameter may be a natural consequence of a low metallicity environment.
The shape of the ionizing radiation field is a strong function of time since the most recent burst of star formation.  The shape of the ionizing radiation field is also determined by the fraction of ionizing photons that are either absorbed by the stellar atmospheres and surrounding gas, and conversely, the fraction of ionizing photons that escape the nebula without being absorbed by ISM.
These effects combine to produce a correlation between ionization parameter and metallicity 
\citep{Dopita86,Dopita00,Kewley02}.  \citet{Dopita06} show theoretically that the ionization parameter inversely correlates with the metallicity of an \HII\ region, $Z$, according to $q \propto Z^{-0.8}$.  The reason for this correlation is twofold; \\

(1) At high metallicity, the stellar wind opacity is larger.  This large opacity absorbs a greater fraction of the ionizing photons, reducing the ionization parameter in the surrounding \HII\ region.  

(2) At high metallicity, stellar atmospheres scatter the photons emitted from the photosphere more efficiently.  This scattering efficiently converts luminous energy flux into mechanical energy flux at the base of the stellar wind.  The reduced luminous energy flux available to ionize the surrounding \HII\ region is observed as a lower ionization parameter.

Many high redshift  galaxies have high ionization parameters \citep[$ -2.9 < \log {\cal U} < -2.0$][]{Pettini01,Lemoine03,Maiolino08,Hainline09,Erb10,Richard11,Wuyts12}.    \citet{Nakajima12} compare the ionization parameter and metallicity (as traced by the \OIIIOII\ vs \R23\ line ratios) for a large sample of high redshift galaxies, including Lyman Break, lensed, and Lyman-$\alpha$ emitting galaxies.   Their comparison shows that the majority of high redshift galaxies have similar ionization parameters to local galaxies at the same metallicity.  Thus, the high ionization parameters at high redshift may simply be a natural consequence of a low metallicity environment.  

The effect of ionization parameter on the \NIIHa\ and \OIIIHb\ ratios depends on the metallicity regime.  Figure~\ref{Kappa_BPT} (left panel) shows that at low metallicities (log(\NIIHa)$<-0.9$), a larger ionization parameter will raise the \OIIIHb\ ratio, while the \NIIHa\ ratio remains constant within $\pm 0.2$~dex.  At high metallicities (log(\NIIHa)$>-0.7$, a larger ionization parameter will reduce the \NIIHa\ ratio while \OIIIHb\ remains roughly constant (to within $\pm 0.1$~dex).  This effect is illustrated  in Figure~\ref{BPT_illustration}.   
For the metallicities of most high redshift galaxies \citep[\OH~$ < 8.6$;][]{Erb06,Wuyts12b,Yuan12},  a high ionization parameter will raise the \OIIIHb\ ratio such that high redshift galaxies will lie above the local abundance sequence, as observed.  Therefore, a high ionization parameter can account for part of the large \OIIIHb\ and \NIIHa\ ratios observed in high redshift galaxies.


\subsection{The High Redshift Star-forming Abundance Sequence} \label{high-z_radiation_field}

Although a hard ionizing radiation field is linked with low metallicity, it is unclear whether hard ionizing radiation fields are a feature of high redshift star-forming galaxies.   Figure~\ref{BPT_illustration} shows that the combination of a larger ionization parameter at low metallicity and a higher electron density may mimic the effects of a harder ionizing radiation field.   Rest-frame UV spectroscopy of high metallicity galaxies at $z>2$ is needed to determine whether the large \NIIHa\ and \OIIIHb\ ratios seen in high redshift galaxies are caused by a higher electron density and ionization parameter or the combined effects of a higher electron density, ionization parameter, {\it and} a harder EUV ionizing radiation field.   

Further, selection biases may be limiting the current range of electron densities and ionization parameters observed at high redshift.   The combination of surface brightness dimming and current observational sensitivities limits high redshift samples to galaxies with high emission-line surface brightness.   This limit biases samples towards galaxies with high star formation rates (SFR$>4$\Msun$/yr$), and more intense star formation.  Local starburst galaxies are characterized by higher pressures than normal \HII\ regions \citep{Kewley01b}.   High pressures of $10^6 < {\rm P}/k < 10^7$ implies gas densities of $100-1000$~cm$^{-3}$.  A greater fraction of starburst galaxies in high redshift samples would produce larger \NIIHa\ and \OIIIHb\ ratios when compared with local samples of normal star-forming galaxies.  

Gravitationally lensed samples can probe an order of magnitude fainter in star formation rate \citep[SFR $> 0.4 $\Msun$/yr$;][]{Richard11}.  Consequently, gravitationally lensed samples cover a larger range of \NIIHa\ and \OIIIHb\ ratios than non-lensed samples \citep{Hainline09,Richard11,Wuyts12}.  However, current lensed samples still do not sample the full range of star formation rates seen in the global spectra of local normal star forming galaxies \citep[ SFR $>0.01$~\Msun/yr][]{Kewley02b}.  A statistically significant sample of high redshift galaxies that covers a broad range of metallicity and stellar mass is required to determine the full range of \NIIHa\ and \OIIIHb\ ratios at high redshift.   Until such samples are available, we assume a lower limit and an upper limit to the star forming abundance sequence at high redshift ($z\sim 3$).    

\subsubsection{Lower Limit Abundance Sequence (Normal SF conditions)} 

Our lower limit abundance sequence is calculated assuming that galaxies at a given redshift have the same shape EUV radiation field, the same range of electron densities, and the same relationship between metallicity and ionization parameter as local star-forming galaxies.   In this very conservative scenario, the star-forming abundance sequence is given  by equation~\ref{SDSS_sequence}.     Chemical evolution moves galaxies down along the star-forming galaxy sequence towards smaller \OIIIHb\ ratios and larger \NIIHa\ ratios.  


\subsubsection{Upper Limit Abundance Sequence (Extreme SF conditions)}


We use PEGASE 2 models \citep{Fioc99} to provide the upper limit to the ionizing radiation field at $z=3$.  These models provide the hardest radiation field because the models use the \citet{Clegg92} planetary nebula nucleus (PNN) atmospheres for stars with high effective temperatures ($T > 50,000 $K).  Clearly, the ionizing spectrum of active star-forming galaxies 
is dominated by the emission from the atmospheres of massive stars, not planetary nebulae.  However, we use the PNN atmospheres as a substitute for massive star stellar atmospheres, in the absence of the full suite of stellar atmospheres with the effects of stellar rotation.  The PEGASE 2 models use the Padova tracks \citep{Bressan93} and the OPAL opacities \citep{Iglesias92}.    The application of PEGASE 2 with our Mappings III code is described in detail in \citet{Kewley01a}.   

If we assume that the hard radiation field sets the upper limit at $z=3$ for star-forming galaxies, then the evolution of the star-forming abundance sequence between $0<z<3$ follows

\begin{eqnarray}
\log({\rm [OIII]/H}\beta) & = &  1.1+ 0.03z    \nonumber \\
& & + \frac{0.61}{\log({\rm [NII]/H}\alpha) +0.08 - 0.1833  z}  \label{O3_N2_seq_z}.
\end{eqnarray}

This upper limit is consistent with the limit that we derive from a combination of a larger ionization parameter ($-2.9 < \log {\cal U}$) and larger electron density ($N_e \sim 1000\,cm^{-3}$)
for metallicities below \OH$\sim 8.8$.  According to our models, the metallicity of a galaxy at redshift $z$ on the star-forming abundance sequence given by equation~\ref{O3_N2_seq_z} is 

 \begin{eqnarray}
 12 + \log{\rm O/H} & = &  8.97 + 0.0663 z -  \nonumber \\
  & & \log(O3N2) (0.32 -0.025 z) \label{Z_seq_z}
\end{eqnarray}

where $O3N2={\rm [OIII]/H}\beta / {\rm [NII]/H}\alpha$.  

\begin{figure}[!h]
\epsscale{1.0}
\plotone{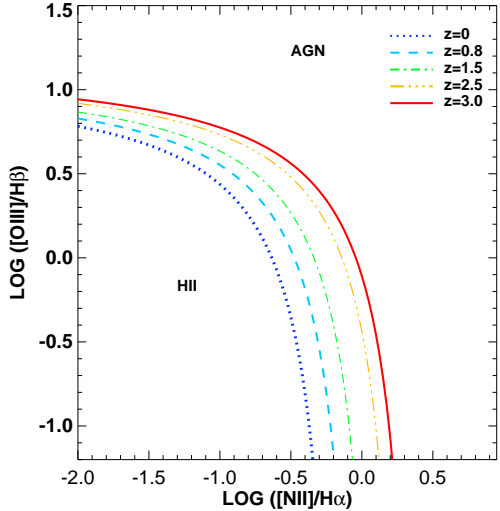}
\caption[BPT_abund_seq_evolution]{The BPT diagram showing the theoretical lower limit ($z=0$; blue dot-dashed line) and upper limit star-forming abundance sequence as a function of redshift ($z=0.8,1.5,2.5,3.0$; dashed, dot-dashed, triple dot-dashed and solid lines, respectively).  
\label{BPT_abund_seq_evolution}}
\end{figure}

In Figure~\ref{BPT_abund_seq_evolution}, we show the evolution of the upper limit star-forming abundance sequence with redshift given by equation~\ref{O3_N2_seq_z}.  Figure~\ref{BPT_abund_seq_evolution} should not be used over small redshift intervals with small samples to test the evolution of the galaxy population because the model errors ($\pm 0.1$~dex) and spread in ionization parameter $log(U)$,  ($\pm 0.1$~dex) at a given redshift are larger than the predicted evolution of the sequence for an ensemble of galaxies across each redshift interval.

\subsubsection{Abundance Sequence Shape}

The metallicity range (or spread) of the star-forming population determines the length of the abundance sequence. 
In cosmological hydrodynamic simulations, the evolution of the lower metallicity bound is dominated by chemical enrichment of the most metal poor galaxies in the galaxy population.   The upper metallicity bound may be influenced by gas inflows from the intergalactic medium such that the upper metallicity bound might fall with redshift \citep{Nagamine01}.   

Again, we assume that at least some galaxies at $z\sim 3$ have reached the level of enrichment of galaxies in the local universe, and we assume that the mean metallicity of the galaxy population evolves according to equation~\ref{eqn_metal_z}.   

The scatter of galaxies about the abundance sequence is dominated by the range in ionization parameters and electron densities in the galaxy population at a given redshift.   How the spread in these properties changes with redshift is unknown.   For simplicity, we assume that the spread in these properties about the mean is constant with redshift  and is $\pm 0.1$~dex.   

\section{The Mixing Sequence} \label{mixing_sequence}

Local active galaxies form two branches on the BPT diagram.  While pure star-forming galaxies lie along the abundance sequence, galaxies with a contribution from a non-thermal radiation field form a sequence that extends towards the pure AGN region of the BPT diagram (i.e. towards large \NIIHa\ and \OIIIHb\ ratios).   This sequence can be produced by either a mixture of gas ionized by hot stars and gas ionized by an AGN \citep{Groves04b}, or a mixture of gas ionized by hot stars and gas ionized by radiative shocks \citep{Kewley01a,Allen08}.  We therefore refer to this sequence as the {\it mixing sequence}, where mixing refers to the mixture of a soft photoionizing radiation field from star-formation and a hard non-thermal radiation field from {\it either} AGN or shocks.  In this section, we focus on starburst-AGN  mixing.  We investigate starburst-shock mixing in Section~\ref{shocks}.

\subsection{AGN Models}

The \NIIHa\ and \OIIIHb\ line ratios of a galaxy purely ionized by an AGN are influenced by the metallicity of the surrounding narrow-line region, the shape of the AGN ionizing radiation field (characterized by a power-law), and the ionization parameter.     We use the Mappings III dusty, radiation pressure dominated models of \citet{Groves04a}.

These models use the same abundance set and depletion factors as our starburst models (Section~\ref{abundance_sequence}).  The ionizing spectrum is based on a power-law,  

\begin{equation}
F_{\mu} \propto \mu^{\alpha}
\end{equation} 

where the frequency $\mu$ is defined over $5 {\rm eV} < \mu < 1000 {\rm eV}$ with four values of the power-law index $\alpha$ ($-1.2,-1.4,-1.7,-2.0$).  

The shape of the ionizing radiation field may change with metallicity; a smaller fraction of metals and dust may change the AGN torus structure and potentially alter the accretion disk. Unfortunately, the effect of metallicity on the AGN ionizing radiation field is unknown.  For simplicity, we assume the same ionizing spectral shape for all metallicities.  

In isobaric radiation-pressure dominated models, the local density varies continually throughout the models.  In this case, the hydrogen density is defined as the density of the \SII\ emission zone, which is located close to the ionization fronts in the Narrow Line Region (NLR).  In this zone, we assume that the electron density is $n_H = 10^3\,{\rm cm}^{-3}$.

The total radiative flux entering the narrow line region cloud is determined by the ionization parameter, $U_{NL}$ defined at the inner edge of the nebula.  The ionization parameter range is $0.0 < \log(U_{NL})< -4.0$ with intervals of $-0.3, -0.6$, and $-0.1$~dex.
The NLR models are truncated at a column density of log[N(H I)] = 21, consistent with observations of NLR clouds \citep{Crenshaw03}.  

\begin{figure}[!h]
\epsscale{0.88}
\plotone{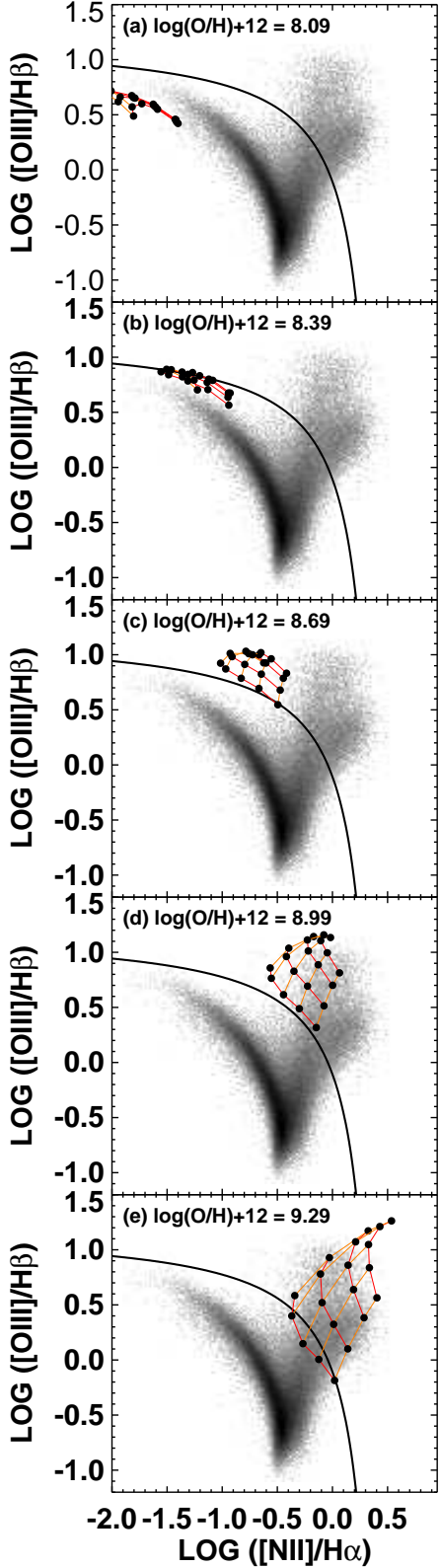}
\caption[AGN_metallicity]{The \citep{Groves04a} dusty radiation pressure dominated photoionization models for AGN as a function of metallicity.  Lines of constant ionization parameter (orange) and constant power-law index (red) are shown.  For comparison, the SDSS sample from \citep{Kewley06a} is shown.  
\label{AGN_metallicity}}
\end{figure}

In Figure~\ref{AGN_metallicity}, we show the position of the dusty radiation-pressure dominated AGN models as a function of metallicity on the BPT diagram.  The position of the AGN models changes substantially with metallicity.  At low metallicity, the AGN models move towards lower \NIIHa\ ratios and the spread in \OIIIHb\ ratios becomes smaller.  The reason for this effect is twofold; (1) at low metallicity, nitrogen shifts from a secondary nucleosynthetic element to a primary nucleosynthetic element, and (2) the rise in electron temperature offsets the fall in oxygen abundance, yielding a roughly constant mean \OIIIHb\ ratio as a function of metallicity.

Figure~\ref{AGN_metallicity} indicates that it will be impossible to distinguish low metallicity galaxies containing AGN (i.e. metallicities \OH$\lesssim 8.4$) from low-metallicity star-forming galaxies using the BPT diagram.

A sample of AGN galaxies at a given NLR metallicity may span a range of ionization parameter and power-law indices.  We therefore use the full suite of ionization parameter and power-law indices to define our 100\% AGN region on the BPT diagram.    We have tested these photoionization models for local galaxies containing AGN; the models produce remarkable agreement with the observed position of AGN on the BPT diagram  \citep{Allen98,Groves04b,Kewley06a}.

\subsection{AGN Metallicity}

Figure~\ref{AGN_metallicity} indicates that AGN reside in metal-rich gas in the local universe ($ 9.0 < \log({\rm O/H})+12 < 9.3 $; KD02 scale).  Indeed, low metallicity AGN are extremely rare in local galaxies \citep{Groves06}.   The star forming abundance sequence and the AGN region on the BPT diagram are linked together by a ``mixing sequence" which can be modelled by a combination of  starburst and power-law AGN photoionization models \citep[e.g.,][]{Groves04b}.  The overlap region between the star-forming abundance sequence and the AGN mixing sequence provides an additional constraint on the metallicity of local galaxies containing AGN.  The local SDSS mixing sequence extends from only the most metal-rich star-forming galaxies ($ 9.0 < \log({\rm O/H})+12 < 9.2 $) to the AGN region.  It is not clear whether AGN are only found in the most metal rich galaxies at higher redshift.   

The combination of metallicity gradients and surface brightness dimming may affect the observed metallicity of galaxies along the mixing sequence.   
The narrow-line region surrounding an AGN is typically 1-5~kpc \citep{Bennert06}, smaller than the typical radius sampled by global spectra of local pure star-forming galaxies \citep[$\sim 8$ kpc on average; ][]{Kewley04}.  Therefore, in galaxies with very steep metallicity gradients, the AGN narrow-line region may be more enriched than the extended star-forming regions which probe a larger radius, on average.  For example, a steep metallicity gradient ($-0.15$~dex~kpc$^{-1}$) can produce a 0.2~dex metallicity difference from spectra measured within circular apertures of 1~kpc and 5~kpc.  On the other hand, a flat metallicity gradient would give the same metallicity for the star-forming gas and the NLR, regardless of the size of the star-forming and narrow-line regions.   Flat metallicity gradients can be produced by large-scale gas flows triggered by galaxy interactions \citep{Kewley06b,Kewley10,Rupke10b,Rich12}.  Theory indicates that metallicity gradients systematically flatten after first pericenter and flatten again during final coalescence \citep{Rupke10,Torrey12}.   The metallicity gradient can be steepened by a late nuclear starburst, providing that outflows have not already removed a substantial fraction of the nuclear star-forming gas (Torrey et al. in prep).

We note that local disk galaxies with typical gradients \citep[$-0.04 \pm 0.09$~dex~kpc$^{-1}$][]{Zaritsky94,vanZee98,Rupke10} are unlikely to produce a large difference in the NLR and star forming regions.  Local early-type galaxies have even shallower gradients \citep{Henry99}. 

Current observations of metallicity gradients at high redshift are limited by small numbers and low angular resolution observations.   Metallicity gradients in some high redshift lensed isolated star-forming galaxies appear to be significantly steeper than local galaxies \citep{Yuan11,Jones13}, but in other high redshift galaxies, the gradients appear to be flat or inverted \citep{Cresci10,Swinbank12,Queyrel12}.   The steepest gradients were obtained by observing gravitationally lensed galaxies with integral field spectrographs with laser-guided adaptive optics \citep{Jones10,Yuan11,Jones13}.  Observational issues such as poor angular resolution ($>0.1" FWHM$) or low S/N can lead to weak line smearing which systematically flattens or inverts metallicity gradients that are intrinsically steep \citep{Yuan13b}.  In the absence of robust observational or theoretical constraints on the cosmic evolution of the metallicity gradient in galaxies, we consider two extreme cases:

\begin{itemize}
\item {\bf Case 1: Metal-rich NLR} The AGN narrow line region at $z=3$ has reached the level of enrichment seen in the narrow line region of local galaxies.  This extreme case implies that galaxies at high redshift have a steeper metallicity gradient on average than observed in local galaxies.      In this scenario, the metallicity gradient of galaxies would flatten with time until $z=0$.

\item {\bf Case 2: Metal-poor NLR} The AGN narrow line region at $z=3$ is at the same average metallicity as the surrounding star-forming gas.   This extreme case implies a flat metallicity gradient on average at high redshift.   If the average galaxy metallicity gradient is flat at high redshift, the gradient steepens with time until $z=0$.  .

\end{itemize}

We interpolate between the model grids in Figure~\ref{AGN_metallicity} to derive the range of possible \NIIHa\ and \OIIIHb\ ratios for the 100\% AGN region at each redshift. The AGN model metallicity is simply the local average metallicity (\OH$\sim 9.0$) for Case 1.  For Case 2, the average AGN model metallicity is determined by the redshift through equation~\ref{eqn_metal_z}.   The spread of the 100\% AGN region on the BPT diagram for both cases is determined by the full range of ionization parameters and power-law indices in our model grid at a given AGN model metallicity.   

We derive starburst-AGN mixing sequences for each redshift using our AGN and starburst models. For both cases, we assume that AGN reside in the most metal-rich galaxies at any epoch, as observed locally.  We assume that the fraction of the starburst sequence that intersects the AGN mixing sequence covers the top $0.2$~dex in metallicity.  If this assumptionÊdoes not hold at high redshift, then we would expect to see galaxies to the left of our mixing sequence (i.e. towards lower \NIIHa\ line ratios than spanned by the mixing sequence).   We fit the mixing sequences with a 2nd or 3rd order least-square polynomial:

\begin{equation}
y = a + b x + c x^2 + d x^3
\end{equation}


where $y=\log {\rm [OIII]/H}\beta$ and $x=\log {\rm [NII]/H}\alpha$.  The constants $a-d$ are given for each mixing line boundary at each redshift in Table~\ref{mixing_table}, along with the range of \NIIHa\ and \OIIIHb\ values over which the mixing sequences are defined.   The four scenarios tabulated in Table~\ref{mixing_table} are described in detail in Section~\ref{Cosmic_BPT}.     


\begin{deluxetable*}{lll|ll|ll|ll}
\tabletypesize{\scriptsize}
\tablecaption{Mixing Sequence Boundaries\label{mixing_table}\tablenotemark{a}}
\tablehead{  & \multicolumn{2}{c|}{Scenario 1}   &   \multicolumn{2}{c|}{Scenario 2}    &   \multicolumn{2}{c|}{Scenario 3} &    \multicolumn{2}{c}{Scenario 4} }
\startdata 
& \multicolumn{2}{l|}{SF model: normal}  &      \multicolumn{2}{l|}{SF model: normal} &  \multicolumn{2}{l|}{SF model: extreme}  &    \multicolumn{2}{l}{SF model: extreme}        \\
& \multicolumn{2}{l|}{AGN model: Metal-rich NLR}  &      \multicolumn{2}{l|}{AGN model: Metal-poor NLR} &  \multicolumn{2}{l|}{AGN model: Metal-rich NLR}  &    \multicolumn{2}{l}{AGN model: Metal-poor NLR}        \\ \hline \hline

& \multicolumn{2}{c|}{$z =       0$} & \multicolumn{2}{c|}{$z =       0$} & \multicolumn{2}{c|}{$z =       0$} & \multicolumn{2}{c|}{$z =       0$} \\ 
&  lower & upper & lower & upper & lower & upper & lower & upper\\ \hline 
 $a$ &     0.034 &      0.885 &     0.029 &      0.885 & 
    0.031 &      0.917 &     0.029 &      0.917 \\
 $b$ &       1.447 &     -0.792 &       1.340 &     -0.792 & 
      1.441 &     -0.491 &       1.332 &     -0.491 \\
 $c$ &     -0.986 &      -6.712 &     -0.712 &      -6.712 & 
    -0.879 &      -6.090 &     -0.710 &      -6.090 \\
 $d$ &  \ldots & \ldots &       1.472 & \ldots &  \ldots &  \ldots &      1.594 & \\ 
$x_{r}$ &     [-0.45, 0.29] &     [-0.45 ,     -0.12] & 
    [-0.45 ,      0.29] &     [-0.45 ,     -0.12] & 
    [-0.50 ,      0.29] &     [-0.44 ,     -0.12] & 
   [ -0.47 ,      0.29] &     [-0.44 ,     -0.12] \\
$y_{r}$ &     [-0.90 ,      0.38] & [    -0.20 ,      0.91] & [
    -0.90 ,      0.38] & [    -0.20 ,      0.91] & [
    -0.90 ,      0.38] & [   -0.10 ,      0.91] & [
     -1.05 ,      0.38] & [   -0.10 ,      0.91] \\ \hline \hline
& \multicolumn{2}{c|}{$z =       0.8$} & \multicolumn{2}{c|}{$z =       0.8$} & \multicolumn{2}{c|}{$z =       0.8$} & \multicolumn{2}{c}{$z =       0.8$} \\ &  lower & upper & lower & upper & lower & upper & lower & upper\\ \hline 
 $a$ &     0.034 &       1.002 &      0.603 &      -13.734 & 
    0.025 &       1.032 &      0.567 &      -6.277 \\
 $b$ &       1.447 &      0.602 &       1.422 &      -38.844 & 
      1.429 &      0.882 &       1.728 &      -18.664 \\
 $c$ &     -0.986 &      -2.078 &      -1.606 &      -25.672 & 
    -0.693 &      -1.382 &      -2.702 &      -11.999 \\
 $d$ &  \ldots  &  \ldots &       5.072 &  \ldots & \ldots  &  \ldots &      6.479 & \\
 $x_{r}$ & [    -0.45 ,      0.29] & [    -0.54 ,     -0.12] & [
    -0.45 ,      0.30] & [    -0.77 ,     -0.59] & [
    -0.37 ,      0.29] & [    -0.49 ,     -0.12] & [
    -0.34 ,      0.30] & [    -0.77 ,     -0.56] \\
$y_{r}$ & [    -0.90 ,      0.38] & [    0.05 ,      0.91] & [
    -0.90 ,       1.00] & [     0.13 ,       1.00] & [
    -0.60 ,      0.38] & [     0.26 ,      0.91] & [
    -0.85 ,       1.00] & [     0.40 ,       1.00] \\ \hline \hline
& \multicolumn{2}{c|}{$z =       1.5$} & \multicolumn{2}{c|}{$z =       1.5$} & \multicolumn{2}{c|}{$z =       1.5$} & \multicolumn{2}{c}{$z =       1.5$} \\ &  lower & upper & lower & upper & lower & upper & lower & upper\\ \hline 
 $a$ &     0.034 &       1.027 &      0.745 &      -10.585 & 
    0.047 &       1.022 &      0.696 &      -6.293 \\
 $b$ &       1.447 &      0.902 &       1.365 &      -26.300 & 
      1.252 &      0.939 &       1.955 &      -15.695 \\
 $c$ &     -0.986 &     -0.837 &     -0.233 &      -14.970 & 
    -0.262 &      0.155 &      -3.789 &      -8.433 \\
 $d$ &   \ldots  &  \ldots  &       10.3905 &  \ldots  &  \ldots  &  \ldots  &      18.5499 & \\ 
  $x_{r}$ & [    -0.45 ,      0.29] & [    -0.61 ,     -0.12] & [
    -0.45 ,      0.17] & [    -0.88 ,     -0.67] & [
    -0.35 ,      0.29] & [    -0.55 ,     -0.12] & [
    -0.25 ,      0.17] & [    -0.88 ,     -0.71] \\
$y_{r}$ & [    -0.90 ,      0.38] & [     0.16 ,      0.91] & [
    -0.90 ,       1.00] & [     0.25 ,       1.00] & [
    -0.40 ,      0.38] & [     0.55 ,      0.91] & [
    -0.60 ,       1.00] & [     0.60 ,       1.00] \\ \hline \hline
& \multicolumn{2}{c|}{$z =       2.5$} & \multicolumn{2}{c|}{$z =       2.5$} & \multicolumn{2}{c|}{$z =       2.5$} & \multicolumn{2}{c}{$z =       2.5$} \\ &  lower & upper & lower & upper & lower & upper & lower & upper\\ \hline 
 $a$ &     0.034 &       1.033 &      0.879 &      -8.604 & 
     0.163 &      0.958 &      0.834 &      -64.678 \\
 $b$ &       1.447 &      0.995 &       1.801 &      -19.285 & 
     0.722 &      0.415 &       2.479 &      -139.855 \\
 $c$ &     -0.986 &     -0.246 &       4.277 &      -9.706 & 
     0.160 &      0.212 &      -3.271 &      -74.252 \\
 $d$ &  \ldots  &  \ldots &       20.374 &  \ldots &  \ldots &  \ldots &      58.818 & \\ 
 $x_{r}$ & [    -0.45 ,      0.29] & [    -0.67 ,     -0.12] & [
    -0.45 ,     0.07] & [    -0.99 ,     -0.75] & [
    -0.30 ,      0.29] & [    -0.70 ,     -0.12] & [
    -0.17 ,     0.07] & [     -1.01 ,     -0.99] \\
$y_{r}$ & [    -0.90 ,      0.38] & [     0.25 ,      0.91] & [
    -0.90 ,       1.00] & [     0.37 ,       1.00] & [
   -0.05 ,      0.38] & [     0.77 ,      0.91] & [
    -0.20 ,       1.00] & [     0.85 ,       1.00] \\ \hline \hline
& \multicolumn{2}{c|}{$z =       3.0$} & \multicolumn{2}{c|}{$z =       3.0$} & \multicolumn{2}{c|}{$z =       3.0$} & \multicolumn{2}{c}{$z =       3.0$} \\ &  lower & upper & lower & upper & lower & upper & lower & upper\\ \hline 
 $a$ &     0.034 &       1.033 &      0.942 &      -8.212 & 
     0.259 &      0.933 &      0.908 &       2.037 \\
 $b$ &       1.447 &       1.005 &       2.268 &      -17.802 & 
     0.370 &      0.190 &       2.692 &       1.186 \\
 $c$ &     -0.986 &     -0.109 &       7.423 &      -8.621 & 
     0.208 &      0.103 &     -0.839 &      0.171 \\
 $d$ &  \ldots  &  \ldots &       26.057 &  \ldots &  \ldots &  \ldots &      71.964 & \\ 
 $x_{r}$ & [    -0.45 ,      0.29] & [    -0.68 ,     -0.12] & [
    -0.45 ,     0.04] & [     -1.03 ,     -0.78] & [
    -0.28 ,      0.29] & [    -0.80 ,     -0.12] & [
    -0.15 ,     0.04] & [     -1.12 ,      -1.03] \\
$y_{r}$ & [    -0.90 ,      0.38] & [     0.30 ,      0.91] & [
    -0.90 ,       1.00] & [     0.40 ,       1.00] & [
     0.16 ,      0.38] & [     0.85 ,      0.91] & [
    -0.10 ,       1.00] & [     0.92 ,       1.00] \\ \hline
\enddata
\tablenotetext{a}{  Mixing line boundaries extending from the star-forming abundance sequence (0\% AGN) to the AGN region (100\% AGN) on the BPT diagram.  The coefficients $a,b,c,d$ are defined according to $y=a + bx + cx^2 + dx^3$ where $y=\log({\rm [OIII]/H}\beta$) and $x=\log({\rm [NII]/H}\alpha$).  The range of $x$ and $y$ values ($x_{r}$, $y_{r}$) over which these polynomials are valid are shown.  Scenarios (1-4) correspond to the 4 limiting scenarios for our theoretical models, described in Section~\ref{Cosmic_BPT} and shown as columns in Figure~\ref{BPT_evol}.}

\end{deluxetable*}


\section{The Cosmic BPT diagram} \label{Cosmic_BPT}

With our theoretical predictions for the star-forming abundance sequence and the AGN mixing sequence, we can predict how the the BPT diagram might appear at different redshifts.  We have two extreme cases for the star-forming galaxies and two extreme cases for the AGN NLR metallicity, which we briefly summarize below.  

Star forming galaxies at high redshift ($z=3$) may have ISM conditions and/or an ionizing radiation field that are the same as local galaxies ({\it normal ISM conditions}) or that are more extreme than local galaxies ({\it extreme ISM conditions}).  Extreme conditions in star-forming galaxies can be produced by either a larger ionization parameter and a more dense interstellar medium, and/or a harder ionizing radiation field.  

The AGN narrow line region at high redshift may have reached the level of enrichment seen in local galaxies ({\it metal-rich}), or it may evolve similarly to the star-forming gas.  In the latter case, the AGN narrow-line region at high redshift would be more {\it metal-poor} than local AGN narrow-line regions.  

Combining our two sets of extreme cases gives four limiting scenarios for the position of galaxies on the BPT diagram at each redshift:

\begin{itemize}

\item {\bf Scenario 1:} Normal ISM conditions, and metal-rich AGN NLR at high-$z$.

\item {\bf Scenario 2:} Normal ISM conditions, and metal-poor AGN NLR at high-$z$.

\item {\bf Scenario 3:} Extreme ISM conditions, and metal-rich AGN NLR at high-$z$.

\item {\bf Scenario 4:} Extreme ISM conditions, and metal-poor AGN NLR at high-$z$.

\end{itemize}

\begin{figure*}[!h]
\epsscale{1.2}
\plotone{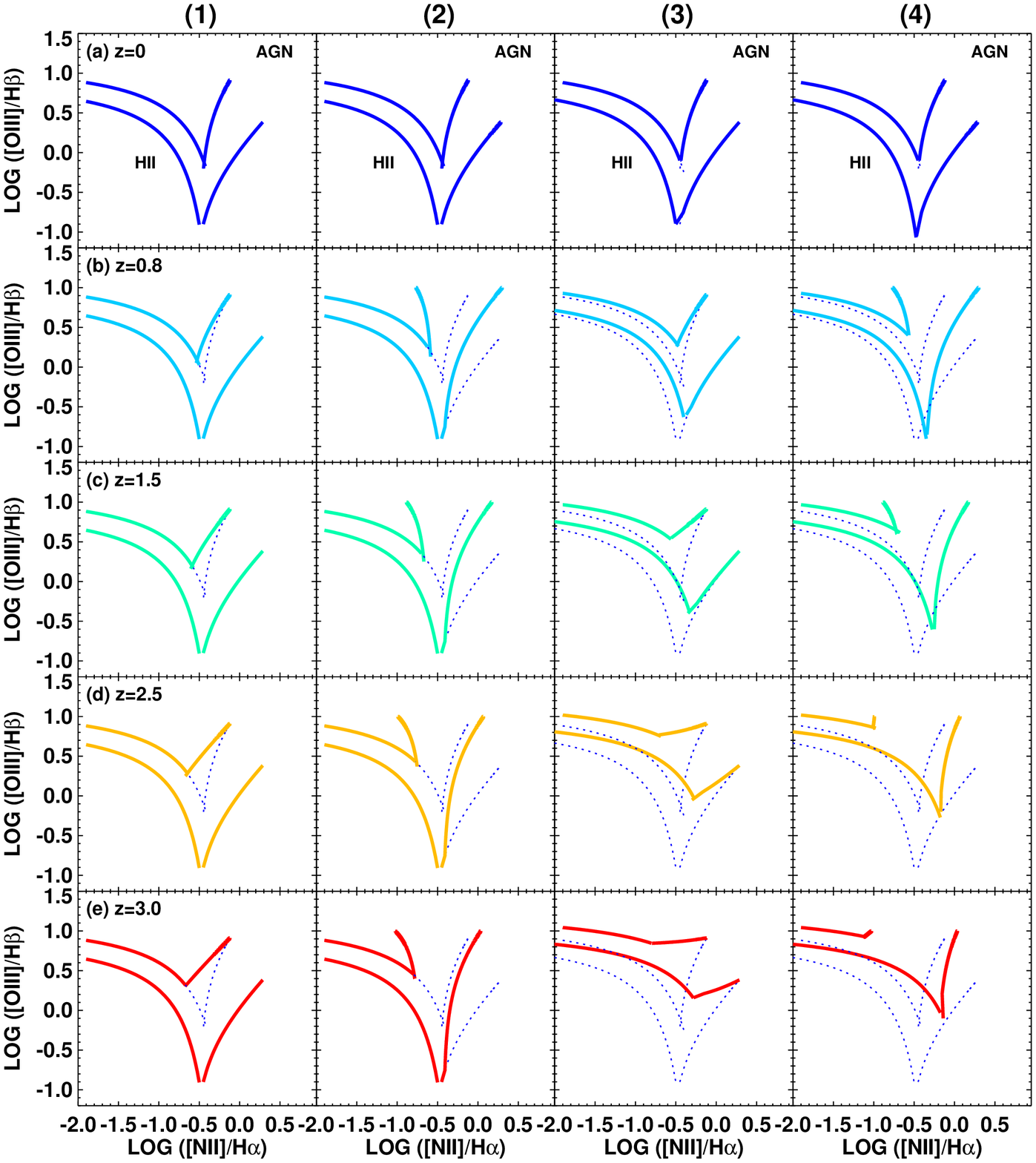}
\caption[BPT_evol]{The Cosmic BPT Diagram: Our theoretical predictions for the position of the star-forming abundance sequence (left curve) and the starburst-AGN mixing sequence (right curve).  The primary driver for our BPT evolution model is chemical enrichment.  The rows give the expected position of the two sequences as a function of redshift for four limiting model scenarios given in each column.  Column (1): Normal ISM conditions, and metal-rich AGN NLR at high-$z$; Column (2): Normal ISM conditions, and metal-poor AGN NLR at high-$z$; Column (3): Extreme ISM conditions, and metal-rich AGN NLR at high-$z$; Column (4): Extreme ISM conditions, and metal-poor AGN NLR at high-$z$.  
\label{BPT_evol}}
\end{figure*}

In Figure~\ref{BPT_evol}, we show how the abundance sequence and the mixing sequence are expected to evolve with redshift given our four limiting scenarios (columns 1-4 in Figure~\ref{BPT_evol}).  The local SDSS sequence boundaries are shown for reference (blue dashed lines).  We discuss the predictions given by each limiting scenario individually below.  

{\it Scenario 1:} In this scenario, we have assumed that the ISM conditions (and/or the ionizing radiation field) are constant as a function of redshift, and the AGN NLR at $z=3$ has already reached the level of enrichment seen in local AGN NLRs.  Therefore, the only change to the position of galaxies on the BPT diagram is a broadening of the mixing sequence at the intersection with the abundance sequence.  This broadening is due to the lower mean and larger range of metallicities in star-forming galaxies at higher redshift.  

{\it Scenario 2:} In this picture, we have assumed that the ISM conditions (and/or the ionizing radiation field) are constant as a function of redshift, and that the AGN NLR metallicity evolves with time in the same way that pure star-forming galaxies evolve (equation~\ref{eqn_metal_z}).  This implies that AGN galaxies at $z=3$ would have flat metallicity gradients that steepen with time.  At high redshift, the mixing sequence occupies lower \NIIHa\ ratios than local galaxies due to the metallicity sensitivity of the \NIIHa\ ratio.  The 100\% AGN region is broad in this scenario, spanning a large range of \NIIHa\ ratios ($-1.0<$\NIIHa$<0$).  The breadth of the AGN region illustrates the effect of differing AGN power-law indices and ionization parameters at lower metallicities.

{\it Scenario 3:} Here, we have assumed that the the ionizing radiation field or the ISM conditions in star forming galaxies become more extreme at high redshift, and that the AGN NLR at $z=3$ has already reached the level of enrichment seen in local AGN NLRs (i.e. a steep metallicity gradient at high-$z$).  In this scenario, the position of the star-forming abundance sequence rises from $z=0$ to $z=3$ towards larger \OIIIHb\ ratios, while the AGN mixing sequence becomes shorter.  At $z=3$, the abundance sequence and AGN mixing sequence form almost a single horizontal sequence across the BPT diagram.  If this scenario exists in high redshift galaxies, classifying galaxies into star-forming and AGN using the BPT diagram would be extremely difficult for $z\geq 2.5$.

{\it Scenario 4:} In this scenario, we have assumed that high redshift star forming galaxies have more extreme ISM conditions and/or a harder ionizing radiation field, and that the AGN NLR metallicity evolves with time in the same way that pure-star-forming galaxies evolve (i.e. a flat metallicity gradient at high redshift).  According to these predictions, the star-forming abundance sequence rises at larger redshift, while the AGN mixing sequence become substantially broader and shorter. At $z=3$, it would be extremely difficult to distinguish between high metallicity star-forming galaxies and galaxies containing AGN using the BPT diagram.

Figure~\ref{BPT_evol} shows that distinguishing between these four scenarios is possible at specific redshifts.  Scenarios (1) and (2) could be easily distinguished with intermediate or high redshift samples ($z\ge 0.8$). All four scenarios   predict substantially different abundance and mixing sequence positions at $z\ge2.5$.  Thus, observations of the rest-frame optical emission-line ratios for statistically significant samples of active galaxies at $z\ge2.5$ are likely to yield important information on the ionizing radiation field  and/or the ISM conditions in star-forming galaxies, as well as the nature of metallicity gradients in galaxies containing AGN at high redshift.

\subsection{The Effect of Shocks} \label{shocks}

Shocks associated with galactic winds may substantially raise the \NIIHa\ emission-line ratio observed in the global spectra of galaxies at high redshift.  Radiative shocks are produced by a variety of astrophysical sources in galaxies, including starburst or AGN-driven outflows, cloud-cloud collisions from galaxy interactions, and jet-cloud collisions.  
In \citet{Rich10} and \citet{Rich11}, we showed that outflows in infrared luminous galaxies can drive gas clouds into the ambient gas in the outer regions of galaxies, creating shock fronts which ionize and excite the gas.  This shocked gas produces strong emission lines at red wavelengths ([NII], [SII], [OI]).  Similar shock excitation has recently been found in  local luminous and ultraluminous infrared galaxies \citep{Sharp10,Soto12,Weistrop12}.   Shocks have also been observed in members of galaxy clusters \citep{Farage10,McDonald12}.    The prevalence of shock excitation in less massive, more normal star-forming systems is unknown.   

Numerous kinematic and detailed spectroscopic studies indicate that galactic winds exist in a significant fraction of galaxies at z$\sim 1$ \citep{Kornei12}, and that winds are prevalent in $z>2$ galaxies \citep[e.g.,][]{Pettini00,Pettini01,Nesvadba07,Steidel10,LeTiran11,Genzel11}.  The fraction of galactic winds that drive shock excitation is unclear.  However, recent integral field spectroscopic studies have revealed the first evidence of shock excitation in high redshift gravitationally lensed galaxies \citep{Yuan12,Newman12}.   Therefore, shock excitation may be an important contributor to the optical line ratios in high redshift galaxies.

Models of photoionized shocks were developed by \citet{Sutherland93} and \citet{Dopita95,Dopita96}, and are described in detail in \citet{Dopita03}.  
We apply the fast shock models of \citet{Allen08} with shock velocities of 300-1000~${\rm km s}^{-1}$ and the slow shock models described in \citet{Rich10,Rich11} covering velocities of 100-200~${\rm km s}^{-1}$.   In slow shocks ($v\lesssim200\,{\rm km s}^{-1}$) the shock front travels faster than the photoionization front emitted by the shocked gas.  In this case, the optical emission-line ratios are directly dependent on the shock velocity.   In fast shocks ($v\gtrsim200\,{\rm km s}^{-1}$) , the ionizing flux emitted by the shock front is large, producing a supersonic photoionization front that moves ahead of the shock front and pre-ionizes the gas.  This photoionization front is referred to as the photoionizing precursor.  The photoionizing precursor can make a significant contribution to the optical emission-lines.  The precursor is usually combined with the theoretical emission from the shocked gas.  For our fast shock models, we assume a 50:50 ratio of shock to precursor emission.  This ratio of shock to precursor emission produces \OIIIHb\ line ratios that are close to the maximum observed in local active galaxies (log(\OIIIHb)$\sim 1.0$).  
We assume a magnetic field strength at pressure equipartition, and a preshock density of $n\sim 1.0\,{\rm cm^{-3}}$.  \citet{Allen08} shows that changing the preshock density from $n\sim 0.01\,{\rm cm^{-3}}$ to $n\sim 1000 \,{\rm cm^{-3}}$ has negligible effect on the position of the equipartition shock+precursor models on the BPT diagram.

\begin{figure}[!h]
\epsscale{0.81}
\plotone{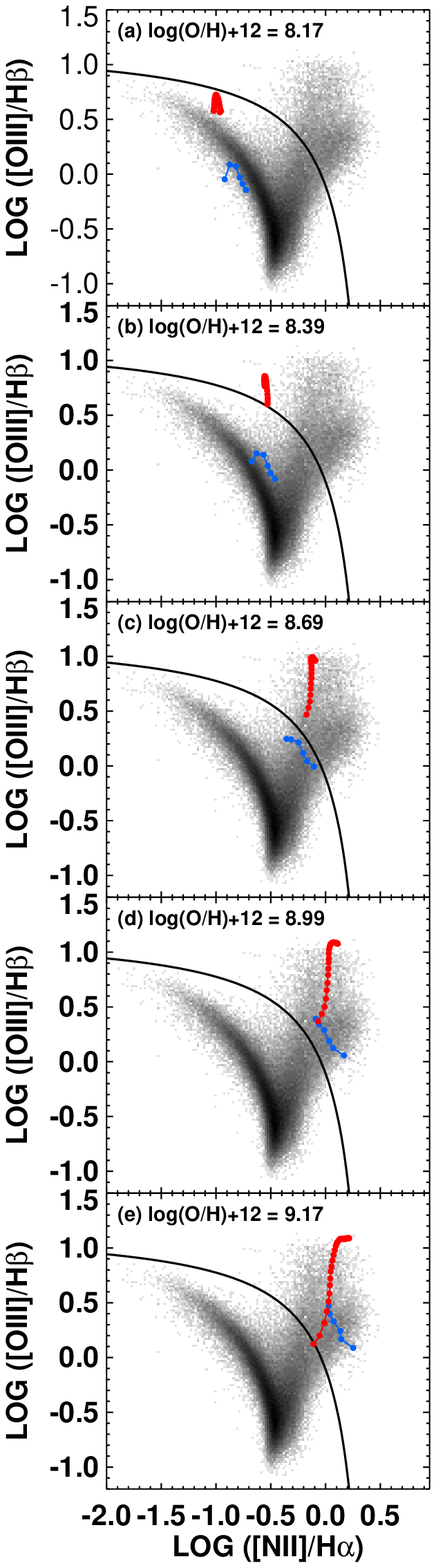}
\caption[Shock_metallicity]{The \citep{Allen08} fast shock models (red) and the \citep{Rich10} slow shock models (blue) as a function of metallicity.  For comparison, the SDSS sample from \citep{Kewley06a} is shown.  The fast and shock slow model positions indicate where galaxies that are 100\% dominated by shocks are likely to lie.  Galaxies containing a mixture of ionizing sources, such as shocks and star-formation will lie along mixing sequences between the star-forming sequence and the 100\% shock models.  The position and shape of the mixing sequences will depend on the metallicity of the galaxy and the shock velocity.  Examples of typical mixing sequences are given in \citet{Rich10,Rich11}.
\label{Shock_metallicity}}
\end{figure}

The full range of shock parameters is explored within the BPT diagram in \citet{Allen08}.   In Figure~\ref{Shock_metallicity}, we show how the position of the fast (red) and slow (blue) shocks change as a function of metallicity and shock velocity.    The colored curves indicate the location of galaxy emission that is fully dominated by shocks.  In reality, shocks rarely dominate a galaxy's global emission.  If a galaxy contains shocks, there may also be a contribution from star-formation and/or AGN.  In the case of composite starburst$+$shock activity, a galaxy will lie along a mixing sequence between the pure star-forming sequence and the 100\% shock models.  
Examples of mixing sequences for nearby galaxies containing both star-formation and shocks are given in \citet{Rich10,Rich11}.

Figure~\ref{Shock_metallicity} shows that the location of the shock models on the BPT diagram is a strong function of metallicity, similar to our AGN models.  The \NIIHa\ ratio becomes 0.3~dex smaller between metallicities typical of local starburst galaxies (\OH$\sim 9.0$) and metallicities typical at high redshift (\OH$\sim 8.7$).  Over this metallicity range, the \OIIIHb\ ratio remains roughly constant within the model errors ($\pm 0.1$~dex) for shocks.  

\begin{figure*}[!h]
\epsscale{1.2}
\plotone{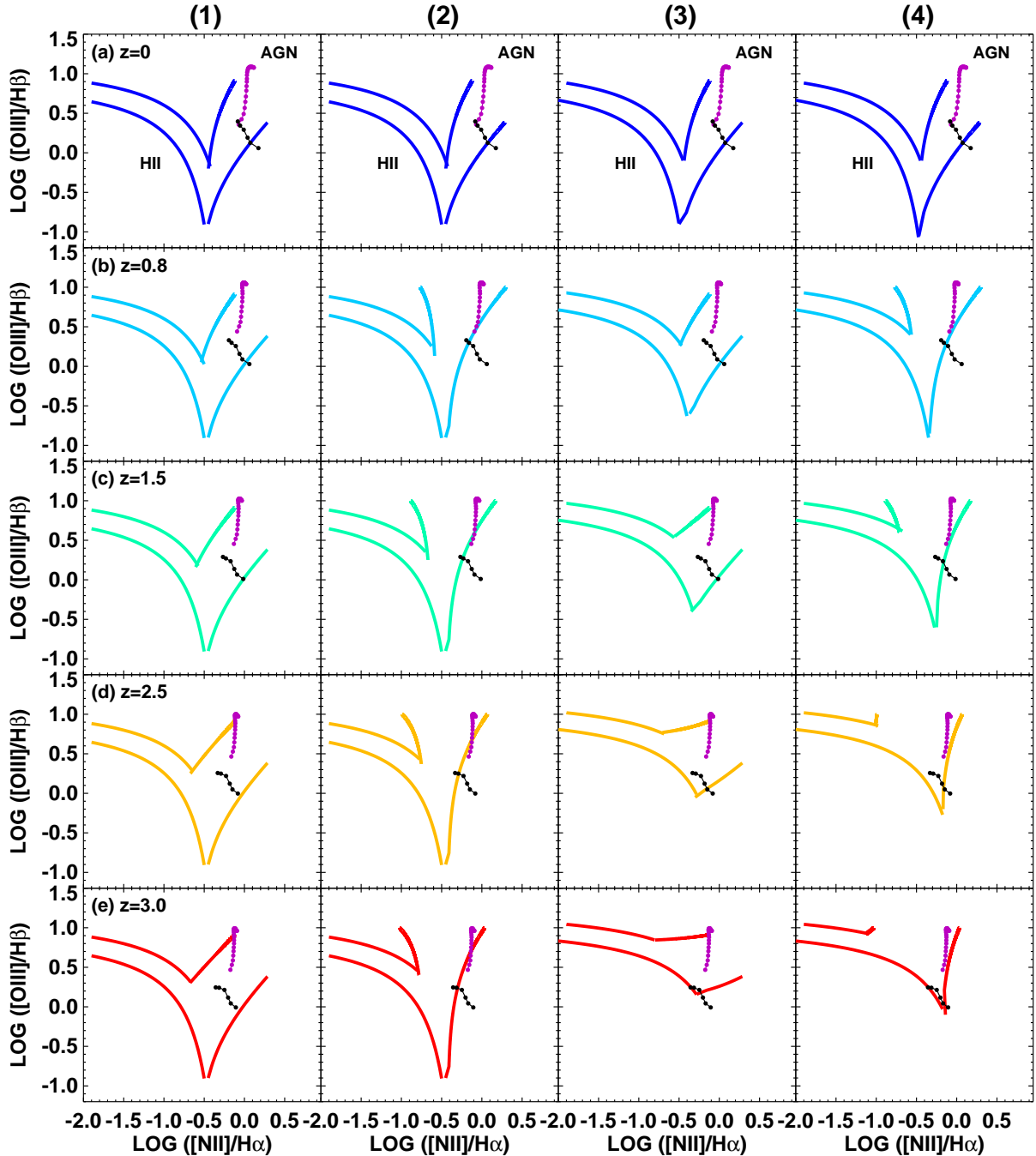}
\caption[Shock_BPT]{The Cosmic BPT Diagram: Our theoretical predictions for the position of the star-forming abundance sequence (left curve) and the starburst-AGN mixing sequence (right curve) as a function of redshift (rows) for our four limiting scenarios (columns) described in Section~\ref{Cosmic_BPT}.   In our model, the primary drivers for BPT evolution are chemical enrichment and a change in the ISM conditions for star-forming galaxies.   The \citep{Allen08} fast shock models (purple) and the \citep{Rich10} slow shock models (black) are shown for comparison.  
\label{Shock_BPT}}
\end{figure*}

Shocks associated with galactic outflows are often spatially distributed throughout the galaxy and are therefore likely to trace regions of similar metallicity to the star forming gas.
 Powerful galactic winds may force metals out of galaxies, while less powerful outflows may act as ``fountains", redistributing metals from the nuclear regions to larger radii \citep[see][for a review]{Putman12}.   Assuming that shocked regions occupy similar metallicities to star-forming regions, we calculate the position of shocks on the cosmic BPT diagram, shown in Figure~\ref{Shock_BPT}.

The position of the 100\% fast and slow shock models on Figure~\ref{Shock_BPT} reveals how a contribution from shock emission (black and purple curves) can mimic non-shock power sources.  Galaxies with a contribution from both star formation and shocks will lie along a mixing sequence between the starburst sequence and the shock models.  Although the exact location of the starburst-shock mixing sequence will depend on the metallicity and on the shock velocity, we can draw two general conclusions from Figure~\ref{Shock_BPT}:

\begin{itemize}

\item Star-forming galaxies with a contribution from fast shocks can mimic a composite starburst-AGN galaxy at all redshifts between $z=0$ to $z=3.0$.   

\item Galaxies containing emission from slow shocks can masquerade as galaxies with composite starburst-AGN activity locally, but may mimic high metallicity starburst galaxies at high redshift ($z>1.5$).   

\end{itemize}

We recommend that galaxies containing (or likely to contain) a significant contribution from shock emission be removed from samples prior to comparison with Figure~\ref{BPT_evol}.   Observations of shock-sensitive optical emission-lines such as \SIIl, \OI, UV shock sensitive lines \citep{Allen98}, or measurements of the velocity dispersion from integral field spectroscopy \citep{Rich11} can help rule out a contribution from shock excitation.  

\subsection{LINERs}

Despite decades of study, the power source of galaxies dominated by emission from Low Ionization Narrow Emission-line Regions (LINERs) is unresolved.  The power source of LINER emission depends critically on sample selection and the aperture used to observe LINER emission.  Optical and X-ray studies suggest that nuclear LINERs ($<1$~kpc) are predominantly inefficiently accreting AGN, analogous to the "low" state of black hole binary systems \citep{Ho05,Satyapal05,Kewley06a,Gonzalez09,Gu09,Ho09}. 
Extended LINER emission ($>1$~kpc) may be produced by shock excitation \citep{Heckman80,Lipari04,Monreal-Ibero06,Sharp10,Rich10,Rich11}. Aged stellar populations in post-starburst galaxies can also produce line ratios typical of LINERs \citep{Taniguchi00,Stasinska08,Yan12,Kehrig12}. 

LINERs lie along the starburst-AGN mixing sequence in the \NIIHa\ versus \OIIIHb\ diagnostic diagram.  LINERs {\it can not} be distinguished from Seyfert galaxies using the  \NIIHa\ versus \OIIIHb\ diagram alone.   Therefore, the mixing sequence on Figure~\ref{BPT_evol} may contain both Seyfert and LINER galaxies, depending on the magnitude limit, sample selection, and power source of the LINER emission.  Thus, the mixing sequence at intermediate and high redshift should not be interpreted as a pure starburst-AGN mixing sequence, unless LINERs and/or shock excitation can be ruled out.


\subsection{Spectral Classification at High Redshift}

In local galaxies, semi-empirical and theoretical classification lines are used to successfully separate star-forming galaxies from those containing an AGN or shock excitation \citep[e.g.,][]{Veilleux87,Kewley01b,Kauffmann03b,Kewley06a}.  These classification lines were developed for and tested on local galaxy samples.  Figure~\ref{BPT_evol} implies that these standard local classification lines should not be applied at high redshift {\it if} the ionizing radiation field is harder, or if the ISM conditions are more extreme (a denser environment or a larger ionization parameter).  Thus, comparisons between Figure~\ref{BPT_evol} and large samples of star-forming galaxies at intermediate and high redshift are critical for testing the applicability of local classification methods to higher redshift samples.  Until such tests have been performed, local classification methods should not be applied to samples with $z \ge 1.5$.

\subsection{Discussion} \label{caveats}

Figure~\ref{BPT_evol} can be used with samples of high redshift galaxies to investigate the ISM conditions in star-forming galaxies at high redshift and, potentially, to constrain metallicity gradients in starburst-AGN composite galaxies as a function of redshift.    For constraining the ISM conditions in star-forming galaxies, redshifts $z>2.5$ is ideal.
For distinguishing between steep  and flat chemical abundance gradients at high redshift, $z=1.5$ is ideal.   At higher redshift, if the ISM conditions in star-forming galaxies are more extreme than locally, distinguishing between steep or flat gradients (i.e. columns 3 and 4) is likely to be difficult, especially with low S/N data.

Our predictions rely on the \citet{Dave11b} cosmological hydrodynamic simulations for the gas-phase metallicity evolution of massive star-forming galaxies.  Initial tests suggest that these simulations successfully reproduce the observed metallicity history for galaxies of intermediate stellar masses ($10^9 M_{\odot} < M_* < 10^{10} M_{\odot}$), but that the simulations predict a larger metallicity for the highest mass galaxies ($10^{10} M_{\odot} < M_*$) than observed for $0<z<3$ \citep{Maiolino08,Yuan12}.  These tests are currently limited by small samples.  If real, we would expect a BPT diagram of the most massive ($10^{10} M_{\odot} < M_*$) galaxies to reveal this difference in two ways:  (1) the observed position at which the mixing sequence and the abundance sequence join would be offset towards larger \OIIIHb\ and smaller \NIIHa\ than the curves shown in Figure~5, and (2) the tip of the AGN sequence may be offset towards lower \NIIHa.  


The best match between a scenario and an ensemble of galaxies may change with redshift.  Galaxies at a given redshift may lie within the boundaries of one scenario, while galaxies at another redshift may lie within (or closer to) the boundaries in another scenario.  Such a change in location with redshift might indicate a more rapid change in properties than we have assumed.  For example, if galaxies at $z=2.5$ fall within the boundaries of Scenario 3 (extreme ISM conditions, steep metallicity gradient), but a statistically significant sample of galaxies at $z=0.8$ fall within the boundaries of Scenario 1 (i.e. local ISM and AGN conditions), then the transition from extreme ISM to local ISM conditions are likely to have occured in the $\sim 4$~Gyr between $0.8<z<2.5$, rather than over the 11~Gyr spanned between $0<z<2.5$.   

Galaxies lying outside our sequence boundaries would provide important constraints for future models of high redshift galaxy spectra.  In general, galaxies lying above the sequence boundaries at a given redshift (i.e. above the abundance sequence and the mixing sequence boundaries in Figure~\ref{BPT_evol}), would indicate that there are either more extreme ISM conditions (or a harder radiation field) than we have assumed in our models, or that some AGN are more metal-poor than we have assumed.  

Note that care must be taken to avoid selection effects when interpreting Figure~\ref{BPT_evol} in terms of the evolution of ensembles galaxies.  For example, local ultraluminous infrared galaxies systematically occupy the mixing sequence and the high metallicity (i.e. large \NIIHa) end of the abundance sequence compared with non-infrared selected galaxies \citep[e.g.,][]{Yuan10}.  A similar selection effect may exist at higher redshift for infrared-selected samples.   Locally, this offset is due to a combination of star-formation, AGN activity, and/or shock excitation \citep{Rich12}.  

Galaxies selected through strong emission-line equivalent widths may also show a bias towards particular locations on the BPT diagram.  Strong \Ha\ equivalent widths are produced by low metallicity galaxies (with little continuum from an old stellar population), and galaxies with new intense bursts of star formation.  Both of these populations may have a higher ionization parameter than in the general population at the same redshift.  These galaxies may lie above the abundance sequence for a volume-limited sample at a given redshift.  

Magnitude-limited samples may preferentially select galaxies with the strongest optical or UV continuum, which may miss the faintest (metal-poor) galaxies and the dustiest (potentially metal-rich) galaxies at a given redshift.   Samples of local galaxies or our theoretical model spectra can be used to investigate which galaxies may be missing from the BPT diagram due to selection biases.  


To summarize, the predictions in Figure~\ref{BPT_evol} may give important insights into both the ISM conditions and metallicity gradients in galaxies at a given redshift, providing that selection effects are well understood in comparison samples.



\section{Conclusions} \label{conclusions}

We have combined current stellar evolutionary synthesis and photoionization models with chemical evolution estimates from cosmological hydrodynamic simulations to predict how the star-forming abundance sequence and the starburst-AGN mixing sequence may appear at intermediate and high redshift on the BPT diagram.   We show that:

\begin{itemize}
\item The position of the star-forming abundance sequence at a given redshift depends on the hardness of the ionizing radiation field and on the ISM conditions of the nebulae surrounding the active star-forming regions.
\item In star-forming galaxies, a harder ionizing radiation field and/or a larger electron density moves galaxies above the normal star-forming abundance sequence on the BPT diagram.  A larger ionization parameter may raise or lower the \NIIHa\ and \OIIIHb\ line ratios, depending on the metallicity of a given galaxy.
\item The position of the AGN branch on the BPT diagram depends sensitively on the metallicity of the AGN narrow-line region.
At low metallicities (\OH$\lesssim 8.4$), the AGN region coincides with the star-forming abundance sequence. Low metallicity AGN galaxies can not be distinguished from low-metallicity star-forming galaxies using the BPT diagram alone.
\item Investigations into the position of the galaxy population on the BPT diagram at a given redshift potentially provides a powerful probe of how the ISM conditions in star forming galaxies and how the metallicity gradient in AGN galaxies has changed with time.
\end{itemize}

We apply our latest fast and slow shock models to investigate how well the BPT diagram can isolate shocked galaxies at intermediate and high redshift.  We conclude that:

\begin{itemize}
\item Galaxies dominated by fast shock emission masquerade as galaxies dominated by AGN at all redshifts.  
\item Galaxies dominated by slow shocks typical of galactic outflows mimic starburst-AGN composites at low redshift, and mimic high metallicity starburst galaxies at
high redshift ($z>1.5$).
\item In galaxies where shocks are suspected, we recommend the use of high resolution integral field spectroscopy to isolate shocked regions from regions of star-formation prior to line ratio analysis.
\end{itemize}

Over the coming decade, near-infrared multi-object spectroscopy and high resolution near-infrared integral field spectroscopy will allow rest-frame optical spectroscopic analysis to be applied to large, statistically significant samples beyond the local universe, for the first time.   These surveys will allow the detailed application of theoretical population synthesis, photoionization, and shock models beyond the local universe, significantly improving our understanding of the ISM conditions in galaxies in the intermediate and high redshift universe.  This work is the first in a series aimed at investigating the ISM conditions in active galaxies across cosmic time.

\acknowledgments
This research has made use of NASA's Astrophysics Data System Bibliographic Services.  L. Kewley wishes to acknowledge the ANU Interstellar Plotters for stimulating discussions and the ANU CHELT Academic Women's Writing Workshop for helping this paper to come to fruition.  L. Kewley gratefully acknowledges support by an Australian Research Council (ARC) Future Fellowship FT110101052 and NSF Early CAREER Award AST0748559.  T.T. Yuan acknowledges a Soroptomist Founder Region Fellowship for Women.  Dopita and Kewley would also like to thank the Australian Research Council for support under Discovery project DP0984657 and Discovery Project DP130104879.

\end{document}